\documentclass[12pt]{article}
\usepackage{graphicx}
\usepackage{hyperref}
\usepackage{url}
\usepackage{listings,jvlisting}
\usepackage{amssymb}
\usepackage{tabularx}
\usepackage{amsmath}
\usepackage{mdframed}
\usepackage{setspace}
\usepackage{here}
\usepackage{authblk}
\usepackage{mathtools}
\usepackage{amsthm}
\usepackage{algorithmic}
\usepackage{algorithm}
\usepackage{cite}
\usepackage{comment}

\newtheorem{theorem}{Theorem}[section]

\usepackage{booktabs, tabularx, makecell, threeparttable, array, adjustbox}
\newcommand{\Nc}{\mathcal{N}}
\newcolumntype{L}[1]{>{\raggedright\arraybackslash}p{#1}}
\newcolumntype{C}{>{\centering\arraybackslash\scriptsize}X}

\def\muh{{\widehat \mu}}

\def\X{{\text{\boldmath $X$}}}

\def\ko{{\overline k}}

\def\Nc{{\cal N}}

\def\Cov{{\rm Cov\,}}
\def\Var{{\rm Var\,}}

\begin{document}
\title{Seamless Phase I--II Cancer Clinical Trials Using Kernel-Based Covariate Similarity}
\author[1]{Kana Makino}
\author[1]{Natsumi Makigusa}
\author[1]{Masahiro Kojima\footnote{Address: 1-13-27 Kasuga, Bunkyo-ku, Tokyo 112-8551, Japan. Tel: +81-(0)3-3817-1949 \quad
E-mail: mkojima263@g.chuo-u.ac.jp}}
\affil[1]{Department of Data Science for Business Innovation, Chuo University}

\maketitle

\begin{abstract}\noindent
In response to the U.S.\ Food and Drug Administration’s (FDA) Project Optimus, a paradigm shift is underway in the design of early-phase oncology trials. To accelerate drug development, seamless Phase I/II designs have gained increasing attention, along with growing interest in the efficient reuse of Phase I data. We propose a nonparametric information-borrowing method that adaptively discounts Phase I observations according to the similarity of covariate distributions between Phase I and Phase II. Similarity is quantified using a kernel-based maximum mean discrepancy (MMD) and transformed into a dose-specific weight incorporated into a power-prior framework for Phase II efficacy evaluation, such as for the objective response rate (ORR). Considering the small sample sizes typical of early-phase oncology studies, we analytically derive a confidence interval for the weight, enabling assessment of borrowing precision without resampling procedures. Simulation studies under four toxicity scenarios and five baseline-covariate settings showed that the proposed method improved the probability that the lower bound of the 95\% credible interval for ORR exceeded a prespecified threshold at efficacious doses, while avoiding false threshold crossings at weakly efficacious doses. A case study based on a metastatic pancreatic ductal adenocarcinoma trial illustrates the resulting borrowing weights and posterior estimates.
\end{abstract}

\par\vspace{4mm}
{\it Keywords: Bayesian optimal interval design; Project Optimus; maximum mean discrepancy}

\section{Introduction}
In response to the U.S.\ Food and Drug Administration’s (FDA) Project Optimus~\cite{fda2023optimus}, a paradigm shift in early-phase trial design is underway in oncology clinical trials. In particular, the previous standard approach of identifying the maximum tolerated dose (MTD) in Phase I and directly using it as the recommended Phase II dose or the recommended Phase III dose—a practice especially common for cytotoxic agents—is being reconsidered. Instead, it is now required to evaluate the efficacy of multiple dose levels in Phase II. Moreover, compared with other therapeutic areas, oncology drug development is generally associated with higher development costs, underscoring the need to optimize trial designs to reduce the overall development burden~\cite{sertkaya2024costs}.

Several Phase I/II designs have been proposed that incorporate both safety and efficacy information for dose-finding without randomization in Phase II. Notable examples include Eff-Tox~\cite{thall2004dose,jin2014using}, BOIN12~\cite{lin2020boin12}, BOIN-ET~\cite{takeda2018boin}, uTPI~\cite{shi2021utpi}, and UNITED~\cite{li2025united}. In contrast, some designs implement randomization in Phase II but use patient data from Phase I solely for selecting doses to be carried forward, without incorporating them into the efficacy analysis of Phase II. Examples of such designs include the method by Hoering et al.~\cite{hoering2011seamless}, DROID~\cite{guo2023droid}, and U-BOIN~\cite{zhou2019utility}.

In recent years, there has been growing interest in leveraging historical data to improve the efficiency of clinical trials. The power prior framework, originally introduced by Ibrahim and Chen (2000)~\cite{ibrahim2000power} and further developed by Ibrahim et al. (2003)~\cite{ibrahim2003optimality}, provides a Bayesian mechanism for incorporating historical information into current analyses. In this framework, the likelihood of the historical data is raised to a power parameter between 0 and 1, allowing for a flexible degree of information borrowing depending on the relevance of the historical data to the current trial. Since its introduction, the power prior has inspired a variety of extensions to address its limitations and adapt it to more complex settings. For example, commensurate priors~\cite{hobbs2011hierarchical} model the relationship between current and historical parameters hierarchically, allowing for partial information borrowing based on their similarity. Bayesian hierarchical models have also been employed to integrate multiple data sources or subgroups~\cite{thall2003hierarchical,berry2013bayesian,chu2018bayesian}. More recently, elastic priors~\cite{jiang2023elastic} have been proposed to dynamically adjust the amount of borrowing based on the level of agreement between the historical and current data. Other works have combined the power prior with propensity score methods to adjust for population differences and improve robustness in real-world applications. For instance, the propensity score–integrated power prior (PSPP) approach~\cite{wang2019propensity} allows for borrowing from external or real-world data while accounting for covariate imbalance, and subsequent extensions have generalized this framework to augment both trial arms~\cite{li2022augmenting} or to incorporate multiple external datasets~\cite{lin2019propensity}.

In parallel, seamless Phase I/II designs have been proposed in which data from Phase I patients are carried forward into randomized Phase II. Zhao et al.~\cite{zhao2024bard} proposed an approach in which patients treated at dose levels selected for Phase II are reused from Phase I, provided that they meet the Phase II eligibility criteria. To account for potential differences in baseline distributions across dose levels, they employed covariate-adaptive randomization (minimization) in Phase II to ensure balance across treatment groups. Similarly, Kitabayashi et al.~\cite{kitabayashi2025boin} proposed a method for seamless Phase I/II designs that borrows Phase I data using the Multi-source Exchangeability Model, allowing information sharing across doses and stages. However, when Phase I is conducted as a first-in-human study, patients may have more advanced disease or worse general condition compared with those enrolled in Phase II, leading to reduced responsiveness to treatment. In such cases, naively incorporating Phase I data into Phase II analyses may reduce the probability of success in Phase II.

This paper proposes an information-borrowing method for cases where the distributions of patient backgrounds differ between Phase I and Phase II. Given the typically small sample sizes in Phase I oncology trials, we avoid model-based borrowing and instead propose a kernel-based approach. To control the extent of borrowing, we define a kernel-based weight taking values in $[0,1]$ and apply it to the power-prior framework, allowing flexible discounting of Phase I data based on similarity. Because small sample sizes may raise concerns about the precision of the estimated weights, we also derive an analytical expression for their confidence intervals, enabling evaluation of the uncertainty associated with the borrowing strength. Furthermore, we assess the performance of the proposed method through extensive simulation studies and demonstrate its practical utility via a case study that illustrates the estimated borrowing weights and corresponding posterior results.

This paper is organized as follows. Section~2 introduces the proposed methodology. Section~3 describes the simulation settings and presents the simulation results. Section~4 provides a case study and its results. Section~5 discusses the findings and implications of the proposed method.

\section{Method}
In the context of seamless Phase I/II cancer trial design, various Phase I designs proposed in the literature can be adopted; there is no requirement to employ a specific design. However, to reliably identify the optimal biological dose (OBD), it may be advantageous to use a design that considers both safety and efficacy in dose escalation and allows the use of backfill cohorts—such as the BF-BOIN-ET design. In this study, we adopt the BF-BOIN-ET design for the Phase I part and conduct a randomized comparison of two dose levels in the Phase II part. In the BF-BOIN-ET design, dose escalation and de-escalation take into account both safety and efficacy, with the aim of identifying not only the maximum tolerated dose (MTD) but also the OBD using a utility score. Let the number of dose levels in the Phase I part be denoted by $J$, and let $n^E_{1j}$ denote the number of patients at dose level $j$ who exhibited an efficacy response. For the Phase II part, the two dose levels are selected as the dose with the highest utility score in the Phase I part and the dose with the second-highest utility score. If there is overlap (i.e., if the same dose is selected twice), the higher dose level is prioritized.

\medskip
\noindent{\it Remark}. If the MTD is the lowest dose and there is only one candidate for the OBD, it is not possible to prepare two dose levels for the Phase II part. Such cases are outside the scope of this study. Typically, the lowest dose is set sufficiently below the range of doses expected to be efficacious, making it unlikely to demonstrate adequate efficacy. Depending on the assumptions made in the Phase I part, such scenarios may lead to early termination or trial failure.
\medskip

In Phase II, we assume that multiple doses are selected from among the safe doses identified in Phase I. Let $J^\ast$ denote the number of selected doses, and assume that patients are randomly assigned to $J^\ast$ treatment arms. Efficacy in each arm is measured by the number of responders, denoted as $n^E_{2j^\ast}$ for the $j^\ast$th group. The primary analysis in Phase II is to determine whether the objective response rate (ORR) in each arm exceeds a predefined threshold $\tau$.

Information from the Phase I data is incorporated into the Phase II analysis using a power prior~\cite{ibrahim2015power}, which down-weights historical data through a discounting parameter. The posterior distribution of the ORR $p$ for dose $j$ in Phase II, borrowing information from Phase I via the power prior, is
\begin{align}
\pi(p \mid n_{1j}, n_{1j}^E, n_{2j}, n_{2j}^E, w_j)
\propto L(p \mid n_{2j}, n_{2j}^E)\,\{L(p \mid n_{1j}, n_{1j}^E)\}^{w_j}\,\pi_0(p),\nonumber
\end{align}
where $L(p \mid n_{2j}, n_{2j}^E)$ and $L(p \mid n_{1j}, n_{1j}^E)$ are binomial likelihoods for Phase II and Phase I, respectively. We assume the prior $\pi_0(p) \sim \mathrm{Beta}(0.05, 0.05)$. The posterior distribution is then
\begin{align}
\mathrm{Beta}\!\left(n_{2j}^E + w_j n_{1j}^E + 0.05,\ (n_{2j}-n_{2j}^E) + w_j(n_{1j}-n_{1j}^E) + 0.05\right).\nonumber
\end{align}

We determine the weight $w_j$ based on differences in covariate distributions between Phase I and Phase II. Let $\X_{1i}$ denote the covariates for the $i$th patient in Phase I, and $\X_{2k}$ denote the covariates for the $k$th patient in Phase II, which may affect the ORR. Because the covariates include both continuous variables (e.g., age) and categorical variables (e.g., sex and tumor stage at diagnosis), we compute distributional similarity via the maximum mean discrepancy (MMD) with a kernel function. Prior to computing the MMD, all continuous covariates are standardized to have mean zero and unit variance using pooled statistics across the Phase I and Phase II datasets, to place covariates on comparable scales and avoid dominance by variables with large ranges. Binary and categorical covariates are converted to indicator variables and left unstandardized, as their scales are inherently bounded. Using these data, the unbiased $U$-statistic version of the MMD is
\begin{align}
MMD_j^2(\X_1,\X_2)
&= \frac{1}{n_{1j}(n_{1j}-1)} \sum_{a=1}^{n_{1j}} \sum_{\substack{b=1\\ b \neq a}}^{n_{1j}} k(\X_{1a}, \X_{1b})
+ \frac{1}{n_{2j}(n_{2j}-1)} \sum_{c=1}^{n_{2j}} \sum_{\substack{d=1\\ d \neq c}}^{n_{2j}} k(\X_{2c}, \X_{2d}) \nonumber\\
&\quad - \frac{2}{n_{1j} n_{2j}} \sum_{a=1}^{n_{1j}} \sum_{c=1}^{n_{2j}} k(\X_{1a}, \X_{2c}).\nonumber
\end{align}
The MMD approaches 0 when the two distributions are similar and increases as the discrepancy grows. Because the maximum value depends on the kernel and may exceed 1, we normalize it to $[0,1]$ and define the weight
\begin{align}
w_j \;=\; 1 - \frac{MMD_j}{MMD_{\max}},\label{eq:wdef}
\end{align}
where $MMD_{\max}$ is the theoretical maximum (e.g., $MMD_{\max}=2$ for a Gaussian kernel). We further derive a confidence interval for $w_j$ as follows.

\begin{theorem}[Confidence Interval for $w_j$]
The $(1-\alpha)\%$ confidence interval is
\begin{align}
\left[w_j - z_{\alpha/2}\frac{\sqrt{V(\X_1,\X_2)}}{MMD_{\max}},\ 
w_j + z_{\alpha/2}\frac{\sqrt{V(\X_1,\X_2)}}{MMD_{\max}}\right],\nonumber
\end{align}
where $z_{\alpha/2}$ is the upper $\alpha/2$ quantile of the standard normal distribution, and
\begin{align}
V(\X_1,\X_2)
&=\frac{4(n_{1j}-2)}{n_{1j}(n_{1j}-1)}\zeta_{1}^\dagger
+\frac{4(n_{2j}-2)}{n_{2j}(n_{2j}-1)}\zeta_{2}^\dagger \nonumber\\
&\quad+\frac{2}{n_{1j}(n_{1j}-1)}\zeta_{1}^{\dagger\dagger}
+\frac{2}{n_{2j}(n_{2j}-1)}\zeta_{2}^{\dagger\dagger}
+\frac{4}{n_{1j}n_{2j}}\zeta_{3}^{\dagger\dagger}+o(n^{-2}),\nonumber
\end{align}
with
\begin{align}
\zeta_{1}^\dagger
&=\Var_{\X_{1a}}\!\left[E_{\X_{1b}}\!\left\{k(\X_{1a},\X_{1b})\right\}\right]
+\Var_{\X_{1a}}\!\left[E_{\X_{2c}}\!\left\{k(\X_{1a},\X_{2c})\right\}\right] \nonumber\\
&\quad-2\Cov_{\X_{1a}}\!\Big(E_{\X_{1b}}\![k(\X_{1a},\X_{1b})],
E_{\X_{2c}}\![k(\X_{1a},\X_{2c})]\Big)=O(1), \label{eq:zeta1dag}\\
\zeta_{2}^\dagger
&=\Var_{\X_{2c}}\!\left[E_{\X_{2d}}\!\left\{k(\X_{2c},\X_{2d})\right\}\right]
+\Var_{\X_{2c}}\!\left[E_{\X_{1b}}\!\left\{k(\X_{1b},\X_{2c})\right\}\right] \nonumber\\
&\quad-2\Cov_{\X_{2c}}\!\Big(E_{\X_{2d}}\![k(\X_{2c},\X_{2d})],
E_{\X_{1b}}\![k(\X_{1b},\X_{2c})]\Big)=O(1),\\
\zeta_{1}^{\dagger\dagger}&=\Var_{\X_{1a},\X_{1b}}[k(\X_{1a},\X_{1b})]=O(1),\\
\zeta_{2}^{\dagger\dagger}&=\Var_{\X_{2c},\X_{2d}}[k(\X_{2c},\X_{2d})]=O(1),\\
\zeta_{3}^{\dagger\dagger}&=\Var_{\X_{1a},\X_{2c}}[k(\X_{1a},\X_{2c})]=O(1). \label{eq:zeta3dag2}
\end{align}
\end{theorem}

The proof, including the derivation from \eqref{eq:zeta1dag} to \eqref{eq:zeta3dag2}, is provided in the Appendix together with plug-in estimator expressions. Assuming $\min(n_{1j},n_{2j})=n\to\infty$, the first-order term is
\[
\frac{4(n_{1j}-2)}{n_{1j}(n_{1j}-1)}\zeta_{1}^\dagger
+\frac{4(n_{2j}-2)}{n_{2j}(n_{2j}-1)}\zeta_{2}^\dagger = O(n^{-1}),
\]
and the second-order term is
\[
\frac{2}{n_{1j}(n_{1j}-1)}\zeta_{1}^{\dagger\dagger}
+\frac{2}{n_{2j}(n_{2j}-1)}\zeta_{2}^{\dagger\dagger}
+\frac{4}{n_{1j}n_{2j}}\zeta_{3}^{\dagger\dagger} = O(n^{-2}).
\]
For the confidence interval, the lower limit was truncated at 0 and the upper limit at 1 if they fell outside $[0,1]$.

In Phase II, dose allocation is randomized. In contrast, in Phase I, patient background distributions may vary across doses; therefore, we computed the MMD separately for each dose. If there is no substantial difference in patient background distributions between doses in Phase I, the MMD may alternatively be computed by pooling across doses.

\section{Simulation}
We conducted Monte Carlo simulations to compare the performance of the proposed information-borrowing method with that of a conventional approach without borrowing. This section first describes the simulation settings and then presents the results.

\subsection{Simulation configuration}
\begin{figure}[htbp]
  \centering
  \includegraphics[width=1\linewidth]{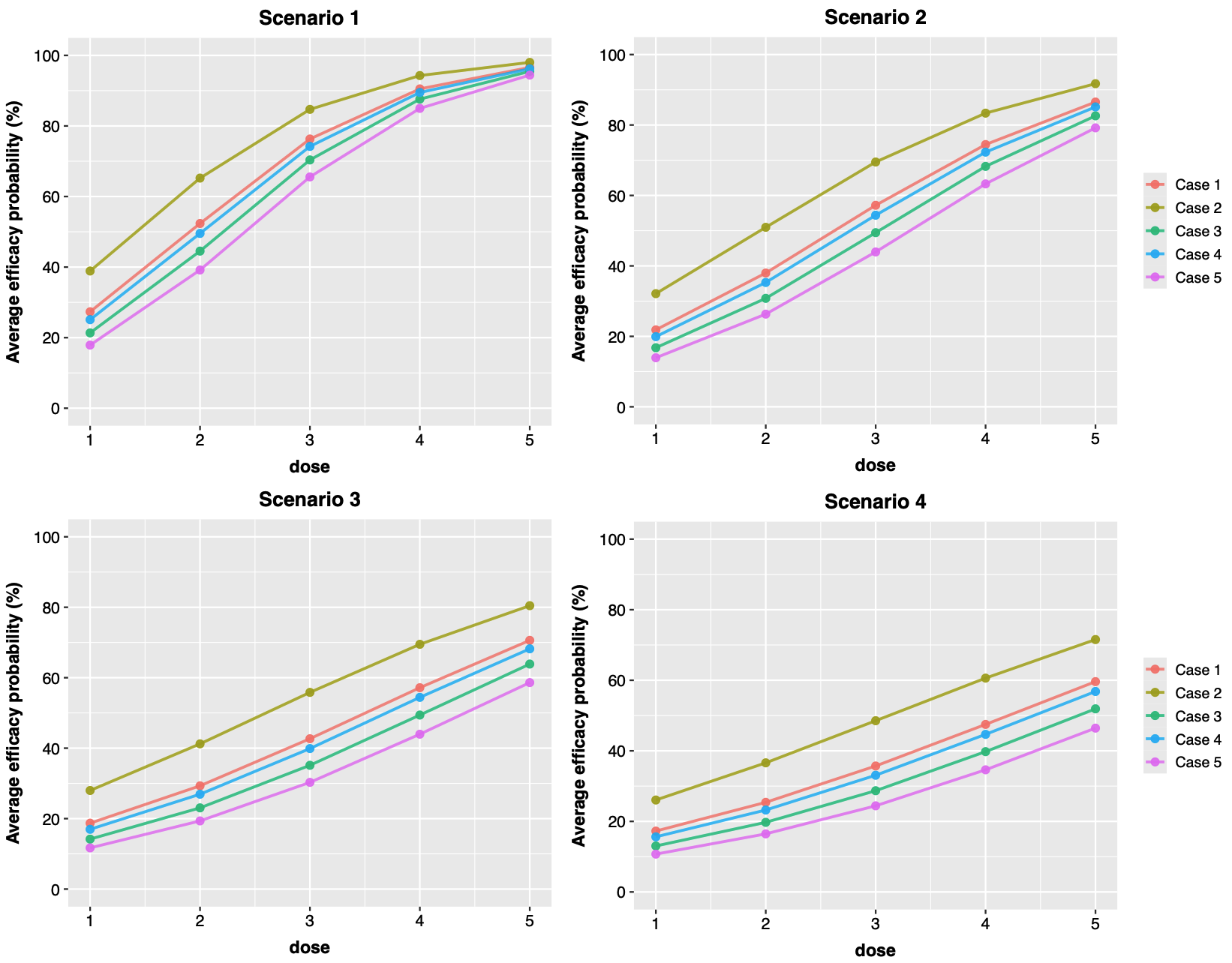}
  \caption{True efficacy probability by dose and case.}
  \label{fig:eff_prob_plot}
\end{figure}

We considered a Phase I/II oncology clinical trial in which the BF-BOIN-ET design was used in Phase I, and a randomized design comparing two dose levels—OBD and OBD--1, both selected from Phase I—was applied in Phase II. The BF-BOIN-ET design was implemented with the following specifications. Five dose levels were considered, with the starting dose set at level 1. Cohorts of three patients were enrolled for a total of ten cohorts in Phase I. The target toxicity and efficacy probabilities were set to 0.30 and 0.25, respectively. For each dose level, the true toxicity probabilities and average efficacy probabilities are summarized in Table~\ref{tab:toxicity_prob} and illustrated in Figure~\ref{fig:eff_prob_plot}. Patient background characteristics were generated according to the values and distributions summarized in the Supplemental Material, specified based on the case study in the next section. Event times for toxicity and efficacy were generated from a Weibull distribution with an assessment window of one month for both endpoints. Patient accrual followed a uniform interarrival distribution with an average rate of three patients per month. Early stopping was triggered when the posterior probability of excessive toxicity or insufficient efficacy exceeded 0.95 and 0.90, respectively. Utility-based OBD selection used pre-specified weights: efficacy without toxicity (100), efficacy with toxicity (60), no efficacy and no toxicity (40), and no efficacy with toxicity (0). Phase II simulations were conducted with a sample size of 20 patients per dose level. Additionally, simulations with a Phase II sample size of 30 per dose level were conducted, and results are provided in the Supplemental Material. For the proposed method, we applied the power prior incorporating the similarity-based weight defined in \eqref{eq:wdef}. As a comparator, we included an approach without information borrowing, relying solely on the Phase II data. The primary endpoint was the probability that the lower bound of the 95\% credible interval for efficacy exceeded 0.3. Results using an alternative threshold of 0.4 are also presented in the Supplemental Material. Simulations were performed under 20 settings, corresponding to combinations of four toxicity scenarios and five baseline-distribution cases. A total of 100{,}000 simulation replicates were performed.

\begin{table}[htbp]
\centering
\caption{Toxicity probabilities for each dose level across scenarios}
\label{tab:toxicity_prob}
\renewcommand{\arraystretch}{1.15}
\setlength{\tabcolsep}{8pt}
\begin{tabular}{lccccc}
\toprule
\textbf{Scenario} & \textbf{Dose 1} & \textbf{Dose 2} & \textbf{Dose 3} & \textbf{Dose 4} & \textbf{Dose 5} \\
\midrule
Scenario 1 & 0.15 & \textbf{0.30} & 0.45 & 0.60 & 0.75 \\
Scenario 2 & 0.05 & 0.15 & \textbf{0.30} & 0.45 & 0.60 \\
Scenario 3 & 0.03 & 0.05 & 0.15 & \textbf{0.30} & 0.45 \\
Scenario 4 & 0.01 & 0.03 & 0.05 & 0.15 & \textbf{0.30} \\
\bottomrule
\end{tabular}
\vspace{0.5em}
\\
\footnotesize Bold values indicate the correct OBD (optimal biological dose).
\end{table}

\begin{table}[htbp]
\centering
\caption{Background information}
\begin{threeparttable}
\renewcommand{\arraystretch}{1.12}
\newcommand{\CatVals}[2]{\mbox{\footnotesize\textit{Cat}\,(#1,\,#2)}}

{\small
\setlength{\tabcolsep}{1.2pt}
\begin{adjustbox}{max width=\linewidth}
\begin{tabularx}{1.15\linewidth}{@{}L{0.19\linewidth} *{5}{>{\centering\arraybackslash\footnotesize}X}@{}}
\toprule
\textbf{Baseline information} & \multicolumn{5}{c}{\textbf{Phase I}} \\[-0.3ex]
\cmidrule(lr){2-6}
& \textbf{Case 1} & \textbf{Case 2} & \textbf{Case 3} & \textbf{Case 4} & \textbf{Case 5} \\
\midrule
Age (years)
  & $\Nc(56.8,\,9.95^{2})$ & $\Nc(45.0,\,10.00^{2})$ & $\Nc(70.0,\,5.00^{2})$ & $\Nc(75.0,\,10.00^{2})$ & $\Nc(80.0,\,5.00^{2})$ \\
Women
  & \textit{Bernoulli}(0.56) & \textit{Bernoulli}(0.90) & \textit{Bernoulli}(0.20) & \textit{Bernoulli}(0.70) & \textit{Bernoulli}(0.20) \\
\makecell[l]{ECOG performance\\status score = 1}
  & \textit{Bernoulli}(0.68) & \textit{Bernoulli}(0.50) & \textit{Bernoulli}(0.80) & \textit{Bernoulli}(0.50) & \textit{Bernoulli}(0.90) \\
\makecell[l]{Tumor stage at\\diagnosis (IIA, III, IV)}
  & \CatVals{0.08}{0.88}
  & \CatVals{0.33}{0.33}
  & \CatVals{0.33}{0.33}
  & \CatVals{0.80}{0.10}
  & \CatVals{0.40}{0.40} \\
\bottomrule
\end{tabularx}
\end{adjustbox}
}

\vspace{0.9em}

{\small
\begin{center}
\begin{minipage}{0.58\linewidth}
\setlength{\tabcolsep}{1.0pt}
\begin{tabularx}{\linewidth}{@{}L{0.50\linewidth} >{\centering\arraybackslash\footnotesize}X@{}}
\toprule
\textbf{Baseline information} & \textbf{Phase II} \\
\midrule
Age (years) & $\Nc(56.8,\,9.95^{2})$ \\
Women & \textit{Bernoulli}(0.56) \\
\makecell[l]{ECOG performance\\status score = 1} & \textit{Bernoulli}(0.68) \\
\makecell[l]{Tumor stage at\\diagnosis (IIA, III, IV)} & \CatVals{0.08}{0.88} \\
\bottomrule
\end{tabularx}
\end{minipage}
\end{center}
}

\vspace{0.3em}
\noindent\footnotesize
\textit{Cat} denotes the \textit{Categorical} distribution. Entries for “Tumor stage at diagnosis (IIA, III, IV)” show the probabilities for (III, IV); the probability for IIA is $1-(\text{III}+\text{IV})$.
\end{threeparttable}
\end{table}

\subsection{Simulation results}
\begin{figure}[htbp]
  \centering
  \includegraphics[width=1\linewidth]{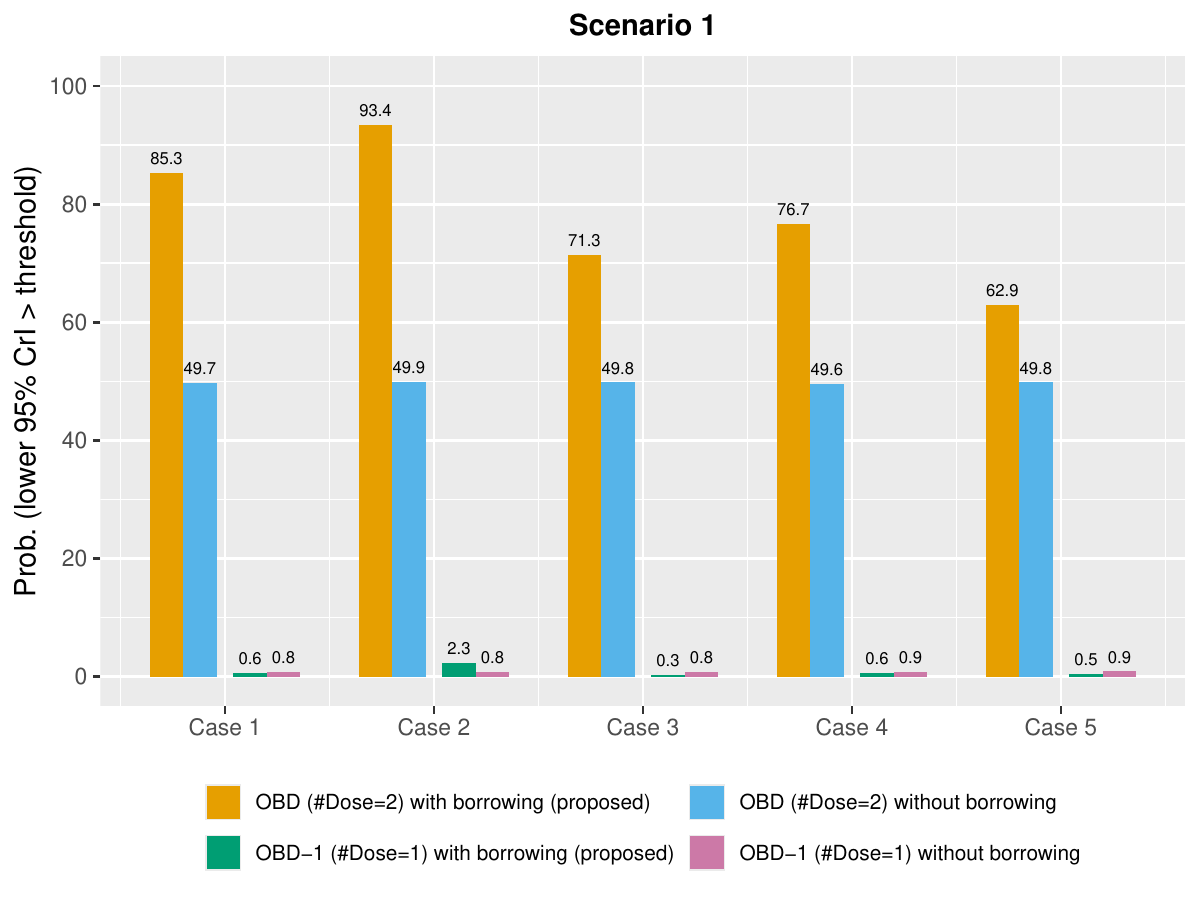}
  \caption{Probability that the lower bound of the 95\% credible interval for efficacy exceeded 0.3 in Scenario 1.}
  \label{fig:method_1_simulation_result_1_CrI_plot}
\end{figure}

\begin{figure}[htbp]
  \centering
  \includegraphics[width=1\linewidth]{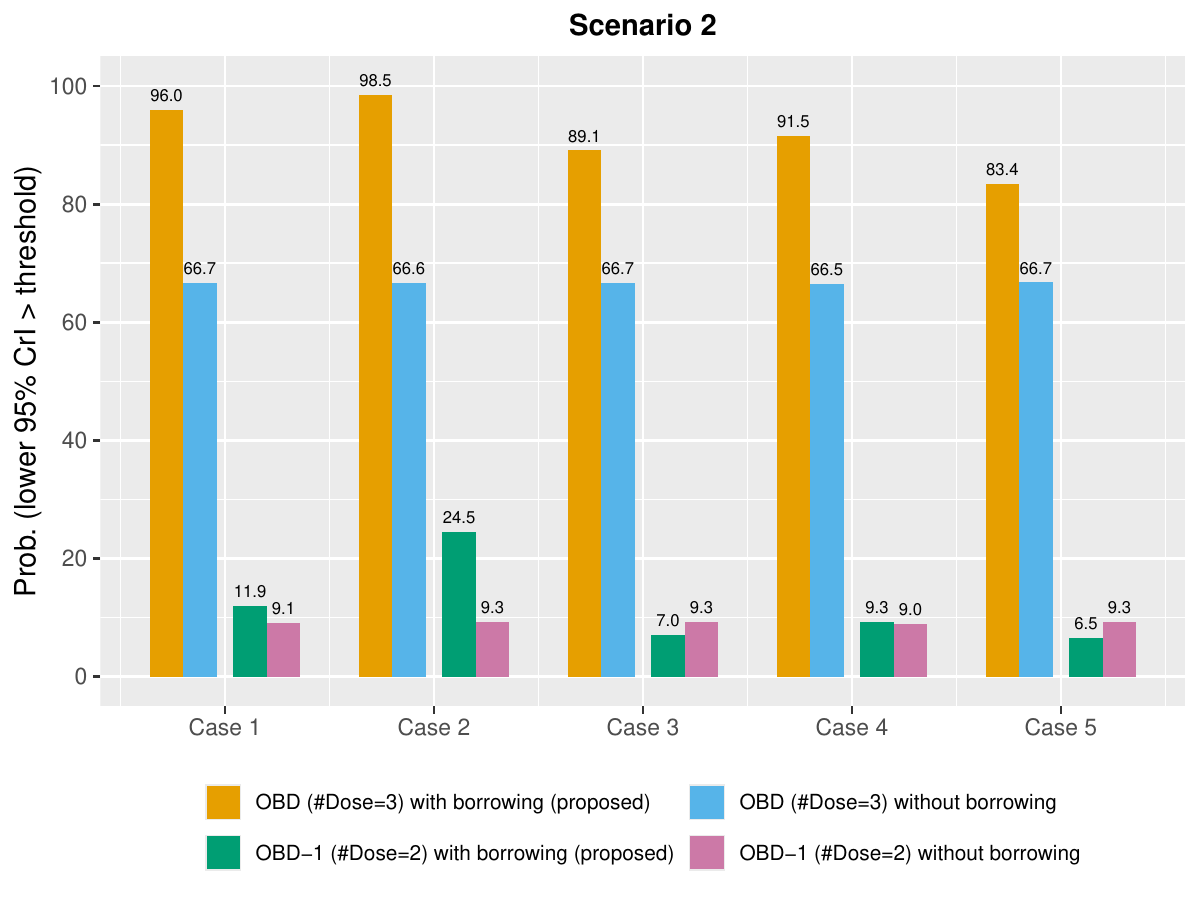}
  \caption{Probability that the lower bound of the 95\% credible interval for efficacy exceeded 0.3 in Scenario 2.}
  \label{fig:method_1_simulation_result_2_CrI_plot}
\end{figure}

\begin{figure}[htbp]
  \centering
  \includegraphics[width=1\linewidth]{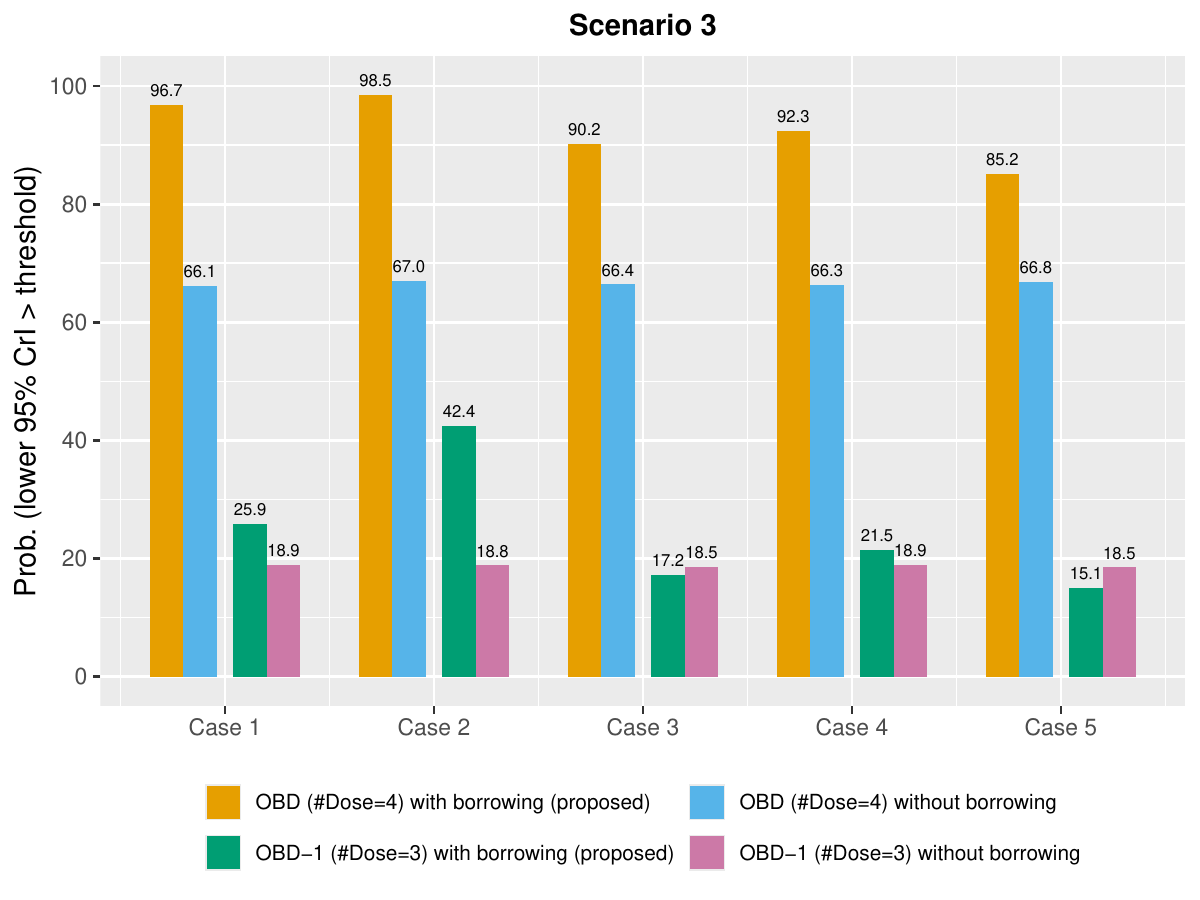}
  \caption{Probability that the lower bound of the 95\% credible interval for efficacy exceeded 0.3 in Scenario 3.}
  \label{fig:method_1_simulation_result_3_CrI_plot}
\end{figure}

\begin{figure}[htbp]
  \centering
  \includegraphics[width=1\linewidth]{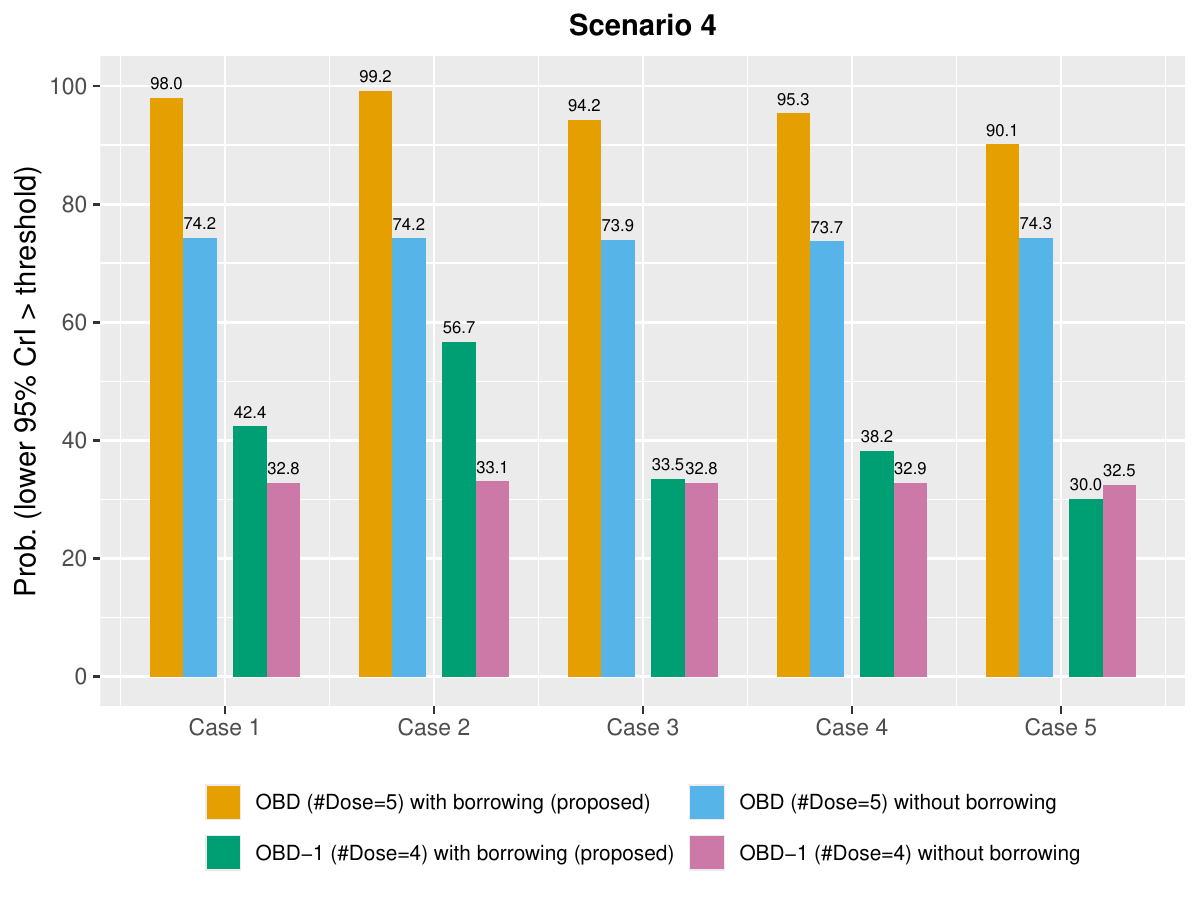}
  \caption{Probability that the lower bound of the 95\% credible interval for efficacy exceeded 0.3 in Scenario 4.}
  \label{fig:method_1_simulation_result_4_CrI_plot}
\end{figure}

This section presents the simulation results for the proposed and comparator methods. The primary outcome was the probability that the lower bound of the 95\% credible interval for efficacy exceeded 0.3. The results are shown in Figures~\ref{fig:method_1_simulation_result_1_CrI_plot}--\ref{fig:method_1_simulation_result_4_CrI_plot}. In addition, detailed results—such as the average number of patients treated per dose level in Phase I, the average observed efficacy probability, and the average borrowing weight—are summarized in tables in the Supplemental Material. Results for the scenario with a Phase II sample size of 30 are also provided in the Supplemental Material.

Across all scenarios, at the true OBD, borrowing increased the probability that the lower bound of the 95\% credible interval for efficacy exceeded 0.3 compared with the analysis without borrowing. For the true OBD--1 dose, when the true efficacy probability was low, borrowing had little impact on this probability. In contrast, when the true efficacy probability was moderately high, borrowing led to a noticeable increase in the probability that the lower bound of the 95\% credible interval for efficacy exceeded 0.3.

\section{Case study}
We conducted a case study based on the Phase I/II trial results for patients with metastatic pancreatic ductal adenocarcinoma (mPDAC) reported by Wainberg et al.~\cite{wainberg2021first} to demonstrate the proposed method. Although only one dose (Dose 2) was selected for Phase II in that trial, we used it as the basis for our case study because background information for each Phase I dose level and the corresponding objective response rate (ORR) results were available. In this case study, we constructed the setting with reference to the trial’s reported background and ORR data. Table~\ref{t:case_ref_summ} summarizes the background information for the hypothetical Phase I and Phase II trials. In Phase I, only the dose levels considered as candidates for Phase II are shown; in practice, data for other dose levels would also have been included. The baseline characteristics considered to potentially influence the ORR are listed in Table~\ref{t:case_ref_summ}.

\begin{table}[H]
\centering
\caption{Reference summary data}
\label{t:case_ref_summ}
\begin{tabular}{|c|c|c|c|c|c|}
\hline
\multicolumn{2}{|c|}{} & \multicolumn{2}{c|}{Phase I} & \multicolumn{2}{c|}{Phase II}\\ \hline
\multicolumn{2}{|c|}{}  & Dose 1 & Dose 2 & Dose 1 & Dose 2\\ \hline
\multicolumn{2}{|c|}{N} & 7 & 7 & 25 & 25\\ \hline
\multicolumn{2}{|c|}{Age (years)} & 66.7 (7.87) & 60.4 (10.66) & 56.8 (9.95) & 56.8 (9.95)\\ \hline
\multicolumn{2}{|c|}{Women} & 6 [85.7] & 4 [57.1] & 14 [56.0] & 14 [56.0]\\ \hline
ECOG performance status score & 0 & 1 [14.3] & 6 [85.7] & 8 [32.0] & 8 [32.0] \\ \hline
ECOG performance status score & 1 & 6 [85.7] & 1 [14.3] & 17 [68.0] & 17 [68.0] \\ \hline
Tumor stage at diagnosis & IIA & 0 & 0 & 1 [4.0] & 1 [4.0] \\ \hline
Tumor stage at diagnosis & III & 3 [42.9] & 1 [14.3] & 2 [8.0] & 2 [8.0] \\ \hline
Tumor stage at diagnosis & IV & 4 [57.1] & 6 [85.7] & 22 [88.0] & 22 [88.0] \\ \hline
\multicolumn{2}{|c|}{Overall response} & 0 & 3 [42.9] & 8 [32.0] & 8 [32.0] \\ \hline
\end{tabular} \\
{\footnotesize Mean (SD). [XX.X] refers to the standard deviation of residual.}
\end{table}

The results of the analysis are presented in Table~\ref{t:case_res}. For Dose 1, the covariate distributions in Phase I and Phase II were moderately similar, yielding a borrowing weight of 0.921. Because no ORR was observed for Dose 1, the estimated ORR was lower than that obtained from Phase II alone. Borrowing information from Phase I resulted in the upper bound of the 95\% credible interval falling below 50\%. For Dose 2, the covariate distributions between Phase I and Phase II were more similar than those for Dose 1, resulting in a higher borrowing weight of 0.970. As the ORR exhibited a consistent pattern in both phases, the posterior mean of the ORR remained largely unchanged; however, borrowing information from Phase I reduced the width of the 95\% credible interval. The 95\% credible interval for the weight was $[0.809, 1.000]$ for Dose 1 ($w_1=0.921$) and $[0.860, 1.000]$ for Dose 2 ($w_2=0.970$).

\begin{table}[H]
\centering
\caption{Posterior estimates of objective response rate (ORR) and borrowing weights in the case study.}
\label{t:case_res}
\begin{tabular}{|c|c|}\hline
Posterior mean [95\% credible interval] & borrowing weight $w$ \\ \hline
\multicolumn{2}{|c|}{Dose 1 in Phase II dynamic borrowing Dose 1 in Phase I} \\ \hline
0.256 [0.122, 0.418] & 0.921 \\ \hline
\multicolumn{2}{|c|}{Dose 1 in Phase II fixed borrowing Dose 1 in Phase I} \\ \hline
0.321 [0.157, 0.511] & 0.000 \\ \hline
0.251 [0.119, 0.412] & 1.000 \\ \hline
\multicolumn{2}{|c|}{Dose 2 in Phase II dynamic borrowing Dose 2 in Phase I} \\ \hline
0.344 [0.192, 0.514] & 0.932 \\ \hline
\multicolumn{2}{|c|}{Dose 2 in Phase II fixed borrowing Dose 2 in Phase I} \\ \hline
0.321 [0.157, 0.511] & 0.000 \\ \hline
0.345 [0.193, 0.514] & 1.000 \\ \hline
\end{tabular} \\
\end{table}

\section{Discussion}
We proposed a nonparametric method to quantify similarity between covariate distributions and to borrow efficacy data from Phase I cancer clinical trials according to the estimated similarity, thereby improving the efficiency of Phase II efficacy evaluation. Because Phase I/II cancer trials usually involve small sample sizes, the precision of the estimated similarity may be limited; hence, we also developed an analytical procedure for constructing its confidence interval.

Although propensity score–based approaches are often used for data borrowing, such methods can be unstable in early-phase cancer trials, where the number of patients per dose level is typically small (3--10 in Phase I and 20--30 in Phase II). Furthermore, in dynamic information-borrowing frameworks such as the power prior, Phase I/II trials are newly conducted studies, and prior reference information for tuning the borrowing parameter is generally unavailable. The proposed method is advantageous because it does not depend on parametric model assumptions and enables information borrowing solely based on the observed data.

Simulation studies were performed to evaluate the performance of the proposed method. The results demonstrated that, for efficacious doses, the proposed information-borrowing approach achieved a higher probability that the lower bound of the 95\% credible interval for efficacy exceeded the predefined threshold compared with the non-borrowing approach. For doses with limited efficacy, borrowing did not lead to falsely exceeding the efficacy threshold proportion. When the Phase I doses had lower efficacy due to background heterogeneity compared with Phase II, the adaptive reduction in the similarity measure effectively mitigated bias, limiting the decrease in this probability to only a few percentage points. By applying backfill cohorts, approximately ten patients on average could be treated in Phase I, which facilitated more efficient information borrowing by stabilizing the estimation of similarity-based weights. With a fixed Phase I sample size, the benefit of information borrowing became more pronounced as the Phase II sample size decreased, suggesting that the proposed approach may be particularly advantageous for sample sizes typical of oncology Phase I/II trials. In each scenario, the borrowing weight in Case 1 was nearly equal to 1. This occurred because the distance between distributions used in the kernel-based similarity measure is invariant; since this invariance does not depend on sample size, the MMD remained close to zero regardless of the number of patients. Because the efficacy probabilities were generated as functions of covariates, type I error rates could not be rigorously evaluated; therefore, simulations assessing type I error were not conducted.

The case study was conducted based on an actual clinical trial for metastatic pancreatic ductal adenocarcinoma (mPDAC). Although slight differences existed in the distributions of patient characteristics, the information-borrowing weight exceeded 0.9. Compared with the analysis without borrowing any information from Phase I, the borrowing approach shifted the estimated efficacy probability toward the overall response observed in Phase I. The lower bounds of the 95\% credible intervals for the weight $w$ exceeded 0.8 for both dose levels, indicating that the covariate distributions between Phase I and Phase II were sufficiently similar to justify information borrowing.

In conclusion, we proposed a nonparametric, kernel-based approach to quantify distributional similarity and incorporated the resulting weight into the power-prior framework to efficiently utilize Phase I data in the evaluation of Phase II trials. Given that early-phase oncology trials typically involve small sample sizes, we also developed an analytical method to construct confidence intervals for the borrowing weights. A limitation of our approach is that if multiple covariates unrelated to efficacy are included and their distributions differ between Phase I and Phase II, the borrowing weight may be unduly reduced. Therefore, careful selection of covariates that are clinically and empirically associated with efficacy is essential when applying the proposed method.

\newpage
\bibliography{main.bib}
\bibliographystyle{unsrt}

\newpage
\begin{appendix}
\section{Appendix}
\label{appa}
\subsection{Proof of Theorem: Derivation of the Variance of MMD}

\begin{align}
MMD_j^2(\X_1,\X_2)
&= \frac{1}{n_{1j}(n_{1j}-1)} \sum^{n_{1j}}_{a=1} \sum^{n_{1j}}_{b \neq a} k(\X_{1a}, \X_{1b})
+ \frac{1}{n_{2j}(n_{2j}-1)} \sum^{n_{2j}}_{c=1} \sum^{n_{2j}}_{d \neq c} k(\X_{2c}, \X_{2d}) \nonumber\\
&\hspace{2em} - \frac{2}{n_{1j}n_{2j}} \sum^{n_{1j}}_{a=1} \sum^{n_{2j}}_{c=1} k(\X_{1a}, \X_{2c}).\nonumber
\end{align}

We decompose the three terms above as follows:
\begin{align}
A_1 &= \frac{1}{n_{1j}(n_{1j}-1)} \sum^{n_{1j}}_{a=1} \sum^{n_{1j}}_{b \neq a} k(\X_{1a}, \X_{1b}),\nonumber\\
A_2 &= \frac{1}{n_{2j}(n_{2j}-1)} \sum^{n_{2j}}_{c=1} \sum^{n_{2j}}_{d \neq c} k(\X_{2c}, \X_{2d}),\nonumber\\
A_{12} &= \frac{2}{n_{1j}n_{2j}} \sum^{n_{1j}}_{a=1} \sum^{n_{2j}}_{c=1} k(\X_{1a}, \X_{2c}),\nonumber
\end{align}
so that
\begin{align}
MMD_j^2(\X_1,\X_2) = A_1 + A_2 - 2A_{12}.\nonumber
\end{align}

Using linearity of variance and the independence between $\X_1$ and $\X_2$, we obtain
\begin{align}
\Var\!\left[MMD_j^2(\X_1,\X_2)\right]
= \Var[A_1] + \Var[A_2] + 4\Var[A_{12}]
- 4\Cov[A_1,A_{12}] - 4\Cov[A_2,A_{12}],\nonumber
\end{align}
where $\Cov[A_1, A_2] = 0$ due to the independence of $\X_1$ and $\X_2$.

\medskip
\noindent\textbf{Variance of $A_1$ and $A_2$.}
Since $A_1$ is a $U$-statistic, we approximate it up to the second-order term using the expansion given in Serfling (2009)~\cite{serfling2009approximation}:
\begin{align}
\Var[A_1]
= \frac{4(n_{1j}-2)}{n_{1j}(n_{1j}-1)} 
\Var_{\X_{1a}}\!\left[E_{\X_{1b}}\!\left\{k(\X_{1a},\X_{1b})\right\}\right]
+ \frac{2}{n_{1j}(n_{1j}-1)} 
\Var_{\X_{1a},\X_{1b}}\!\left[k(\X_{1a},\X_{1b})\right]
+ o(n^{-2}).\nonumber
\end{align}
Similarly, for $A_2$ we have
\begin{align}
\Var[A_2]
= \frac{4(n_{2j}-2)}{n_{2j}(n_{2j}-1)} 
\Var_{\X_{2c}}\!\left[E_{\X_{2d}}\!\left\{k(\X_{2c},\X_{2d})\right\}\right]
+ \frac{2}{n_{2j}(n_{2j}-1)} 
\Var_{\X_{2c},\X_{2d}}\!\left[k(\X_{2c},\X_{2d})\right]
+ o(n^{-2}).\nonumber
\end{align}

\medskip
\noindent\textbf{Variance of $A_{12}$.}
For the cross term $A_{12}$, there are three possible types of index overlap, depending on whether pairs share one or both indices: (i) pairs share the same $\X_1$, yielding $n_{1j}\times n_{2j}(n_{2j}-1)$ combinations; (ii) pairs share the same $\X_2$, yielding $n_{2j}\times n_{1j}(n_{1j}-1)$ combinations; and (iii) identical pairs, corresponding to $n_{1j}n_{2j}$ cases in total. Thus,
\begin{align}
\Var[A_{12}] 
&= \frac{n_{2j}-1}{n_{1j}n_{2j}} 
\Var_{\X_{1a}}\!\left[E_{\X_{2c}}\!\left\{k(\X_{1a},\X_{2c})\right\}\right]
+ \frac{n_{1j}-1}{n_{1j}n_{2j}} 
\Var_{\X_{2c}}\!\left[E_{\X_{1b}}\!\left\{k(\X_{1b},\X_{2c})\right\}\right] \nonumber\\
&\hspace{2em}
+ \frac{1}{n_{1j}n_{2j}} 
\Var_{\X_{1a},\X_{2c}}\!\left[k(\X_{1a},\X_{2c})\right].\nonumber
\end{align}

\medskip
\noindent\textbf{Covariance terms.}
\begin{align}
\Cov[A_1,A_{12}]
&= \frac{2}{n_{1j}}
\Cov_{\X_{1a}}\!\left(
E_{\X_{1b}}\![k(\X_{1a},\X_{1b})],
E_{\X_{2c}}\![k(\X_{1a},\X_{2c})]
\right),\nonumber\\
\Cov[A_2,A_{12}]
&= \frac{2}{n_{2j}}
\Cov_{\X_{2c}}\!\left(
E_{\X_{2d}}\![k(\X_{2c},\X_{2d})],
E_{\X_{1b}}\![k(\X_{1b},\X_{2c})]
\right).\nonumber
\end{align}

Therefore, the expressions from \eqref{eq:zeta1dag} to \eqref{eq:zeta3dag2} are obtained. Each variance and covariance cannot be used directly without substituting the corresponding estimators. Accordingly, the plug-in estimators are defined as follows:
\begin{align}
&\widehat{\Var}_{\X_{1a}}\!\left[\widehat{E}_{\X_{1b}}\!\{k(\X_{1a},\X_{1b})\}\right]
= \frac{1}{n_{1j}-1} \sum_{a=1}^{n_{1j}} 
\left\{\muh_{\X_1|\X_1}(a) - \ko_{\X_1\X_1}\right\}^2, \nonumber\\
&\widehat{\Var}_{\X_{1a}}\!\left[E_{\X_{2c}}\!\{k(\X_{1a},\X_{2c})\}\right]
= \frac{1}{n_{1j}-1} \sum_{a=1}^{n_{1j}} 
\left\{\muh_{\X_2|\X_1}(a) - \ko_{\X_1\X_2}\right\}^2, \nonumber\\
&\widehat{\Cov}_{\X_{1a}}\!\left[
E_{\X_{1b}}\![k(\X_{1a},\X_{1b})],
E_{\X_{2c}}\![k(\X_{1a},\X_{2c})]
\right]
= \frac{1}{n_{1j}-1} \sum_{a=1}^{n_{1j}} 
\left\{\muh_{\X_1|\X_1}(a) - \ko_{\X_1\X_1}\right\}
\left\{\muh_{\X_2|\X_1}(a) - \ko_{\X_1\X_2}\right\}, \nonumber\\
&\widehat{\Var}_{\X_{2c}}\!\left[E_{\X_{2d}}\!\{k(\X_{2c},\X_{2d})\}\right]
= \frac{1}{n_{2j}-1} \sum_{c=1}^{n_{2j}} 
\left\{\muh_{\X_2|\X_2}(c) - \ko_{\X_2\X_2}\right\}^2, \nonumber\\
&\widehat{\Var}_{\X_{2c}}\!\left[E_{\X_{1b}}\!\{k(\X_{1b},\X_{2c})\}\right]
= \frac{1}{n_{2j}-1} \sum_{c=1}^{n_{2j}} 
\left\{\muh_{\X_1|\X_2}(c) - \ko_{\X_1\X_2}\right\}^2, \nonumber\\
&\widehat{\Cov}_{\X_{2c}}\!\left[
E_{\X_{2d}}\![k(\X_{2c},\X_{2d})],
E_{\X_{1b}}\![k(\X_{1b},\X_{2c})]
\right]
= \frac{1}{n_{2j}-1} \sum_{c=1}^{n_{2j}}
\left\{\muh_{\X_2|\X_2}(c) - \ko_{\X_2\X_2}\right\}
\left\{\muh_{\X_1|\X_2}(c) - \ko_{\X_1\X_2}\right\}.\nonumber
\end{align}

For the second-order terms derived from properties of $U$-statistics:
\begin{align}
&\widehat{\Var}_{\X_{1a},\X_{1b}}[k(\X_{1a},\X_{1b})]
= \frac{1}{n_1(n_1-1)} \sum_{a=1}^{n_1} \sum_{b \ne a}
\big\{k(\X_{1a},\X_{1b}) - \bar{k}_{\X_1\X_1}\big\}^2 + o(n^{-2}), \nonumber\\
&\widehat{\Var}_{\X_{2c},\X_{2d}}[k(\X_{2c},\X_{2d})]
= \frac{1}{n_2(n_2-1)} \sum_{c=1}^{n_2} \sum_{d \ne c}
\big\{k(\X_{2c},\X_{2d}) - \bar{k}_{\X_2\X_2}\big\}^2 + o(n^{-2}), \nonumber\\
&\widehat{\Var}_{\X_{1a},\X_{2c}}[k(\X_{1a},\X_{2c})]
= \frac{1}{n_1 n_2} \sum_{a=1}^{n_1} \sum_{c=1}^{n_2}
\big\{k(\X_{1a},\X_{2c}) - \bar{k}_{\X_1\X_2}\big\}^2 + o(n^{-2}).\nonumber
\end{align}

The sample means are
\begin{align}
\ko_{\X_1\X_1}
&= \frac{1}{n_{1j}(n_{1j}-1)} \sum_{a=1}^{n_{1j}} \sum_{\substack{b=1\\ b \neq a}}^{n_{1j}} k(\X_{1a},\X_{1b}),\nonumber\\
\ko_{\X_1\X_2}
&= \frac{1}{n_{1j}n_{2j}} \sum_{a=1}^{n_{1j}} \sum_{c=1}^{n_{2j}} k(\X_{1a},\X_{2c}),\nonumber\\
\ko_{\X_2\X_2}
&= \frac{1}{n_{2j}(n_{2j}-1)} \sum_{c=1}^{n_{2j}} \sum_{\substack{d=1\\ d \neq c}}^{n_{2j}} k(\X_{2c},\X_{2d}),\nonumber\\
\muh_{\X_1|\X_1}(a)
&= \frac{1}{n_{1j}-1} \sum_{\substack{b=1\\ b \neq a}}^{n_{1j}} k(\X_{1a},\X_{1b}),\nonumber\\
\muh_{\X_2|\X_1}(a)
&= \frac{1}{n_{2j}} \sum_{c=1}^{n_{2j}} k(\X_{1a},\X_{2c}),\nonumber\\
\muh_{\X_1|\X_2}(c)
&= \frac{1}{n_{1j}} \sum_{a=1}^{n_{1j}} k(\X_{1a},\X_{2c}),\nonumber\\
\muh_{\X_2|\X_2}(c)
&= \frac{1}{n_{2j}-1} \sum_{\substack{d=1\\ d \ne c}}^{n_{2j}} k(\X_{2c},\X_{2d}).\nonumber
\end{align}

The right-hand sides of 
$\widehat{\Var}_{\X_{1a},\X_{1b}}[k(\X_{1a},\X_{1b})]$,
$\widehat{\Var}_{\X_{2c},\X_{2d}}[k(\X_{2c},\X_{2d})]$, and
$\widehat{\Var}_{\X_{1a},\X_{2c}}[k(\X_{1a},\X_{2c})]$
represent only the first-order terms implied by $U$-statistic theory.
\end{appendix}

\end{document}